\documentclass[aps,prd,amsfonts,amssymb,amsmath,nofootinbib,showpacs,superscriptaddress,onecolumn,floatfix]{revtex4-1}
\usepackage{graphicx}
\usepackage{color}
\usepackage{mathrsfs}
\usepackage[caption=false]{subfig}
\usepackage{dcolumn}
\usepackage[breaklinks=true]{hyperref}
\usepackage{textcomp}
\usepackage{multirow}
\usepackage{epsfig}
\usepackage{graphicx,color,bm}

\newcommand{\T}[1]{\text{#1}}

\newcommand{\rmd}{\mathrm{d}}
\def\be{\begin{equation}}
\def\ee{\end{equation}}
\def\bea{\begin{eqnarray}}
\def\eea{\end{eqnarray}}

\newcommand{\bes}{\begin{subequations}}
\newcommand{\ees}{\end{subequations}}

\allowdisplaybreaks

\begin{document}


\title{
Analytical solutions of bound timelike geodesic orbits in effecitive-one-body frame
}

\author{Chen Zhang}
\affiliation{Shanghai University of Engineering Science, Shanghai, 201620, China}
\author{Wen-Biao Han}
\email{Corresponding author: wbhan@shao.ac.cn}
\affiliation{Shanghai Astronomical Observatory, Shanghai, 200030, China}
\affiliation{School of Fundamental Physics and Mathematical Sciences, Hangzhou Institute for Advanced Study, UCAS, Hangzhou 310024, China}
\affiliation{School of Astronomy and Space Science, University of Chinese Academy of Sciences, Beijing 100049, China}
\affiliation{State Key Laboratory of Radio Astronomy and Technology,
A20 Datun Road, Chaoyang District, Beijing, 100101, China}

\date{\today} 

\begin{abstract}
We derive the approximate analytical solutions of the bound timelike geodesic orbits in the effective-one-body (EOB) frame with extreme-mass ratio limit. The analytical solutions are expressed in terms of the elliptic integrals using Mino time $\lambda$ as the independent variable. Since Mino time decouples the $r$ and $\theta$-motion, we also give explicit expressions for three orbital frequencies $\Omega_r, ~\Omega_\theta, ~\Omega_\phi$ using the Fourier series expansion. With these analytical expressions at hand, we can perform Fourier expansions in Mino time $\lambda$  for any function expressed in terms of the coordinates $(r,\theta,\phi)$. In particular, the observer's time t is decomposed into Mino time $\lambda$, and the frequency-domain description is constructed from the $\lambda$-Fourier expansion and the expansion of t. These analytical expressions are quite simple to implement, and can be applicable for calculating gravitational waves (GWs) from extreme mass-ratio inspirals (EMRIs) with the frequency-domain Teukolsky equation.
\end{abstract}

\maketitle
\section{Introduction}

The successful detection of gravitational waves (GWs) by Advanced LIGO and Virgo~\cite{abbott2016observation,abbott2016gw151226,scientific2017gw170104,abbott2017gw170608,abbott2017gw170814,abbott2017gw170817} has inaugurated the era of GW astronomy. Ground-based detectors observe gravitational waves at high frequencies, while planned space-borne gravitational wave detectors—such as the Laser Interferometer Space Antenna (LISA)~\cite{LISA}, Taiji~\cite{hu2017taiji}, and TianQin~\cite{luo2016tianqin} focus on the low-frequency band (approximately 0.1 mHz to 1 Hz). Extreme-mass-ratio inspirals (EMRIs)~\cite{Kostas_Review}—stellar-mass compact objects (neutron star or stellar-mass black hole) inspiraling into supermassive black holes (SMBH) of $10^5$–$10^7 M_\odot$—represent a prime target for space-based gravitational-wave astronomy. The majority of observed EMRI events are expected to be stellar-mass black hole (SOBH)-SMBH mergers, as mass segregation favors heavier BHs near the galactic centers, and their stronger GW signals enhance detectability. Astrophysical black holes are generally expected to possess spin, which is described within general relativity by the Kerr metric. The secondary BH is often modeled as a test particle, moving on a nearly timelike Kerr geodesic in the SMBH’s spacetime over short orbital time scale. 

The Kerr geodesics have been extensively studied since the discovery of the Kerr solution. They represent a significant subject not only in the mathematical framework of general relativity but also in astrophysical applications. Currently, many black hole candidates have been identified in the universe, spanning a broad range of mass scales, from stellar-mass black holes to those in the nuclei of galaxies. One way to explore the properties of a Kerr black hole spacetime is to study geodesic motion. Detailed works on geodesic motion in black hole spacetimes are comprehensively summarized by Chandrasekhar~\cite{Chandra}. In the weak-field regime, at large distances from the black hole, the orbits of a particle closely resemble those in Newtonian gravity. However, in the strong-field regime, the orbits become significantly more complex, making comparisons with Newtonian gravity challenging. For bound geodesics, this complexity arises due to mismatches between the fundamental frequencies of radial motion, $\Omega_r$, polar motion, $\Omega_\theta$, and azimuthal motion, $\Omega_\phi$. For instance, $\Omega_\phi - \Omega_\theta$ represents the precession of the orbital plane, while $\Omega_\phi - \Omega_r$ describes the precession of the orbital ellipse. These frequency differences grow significantly in the strong-field regime, particularly near the black hole's event horizon or close to the separatrix—the boundary separating stable and unstable orbits.

The complex behavior discussed above demonstrates the crucial role of fundamental frequencies in analyzing bound geodesic orbits. However, the coupling of the $r$ and $\theta$ motions in the geodesic equation has prevented one from deriving the fundamental frequencies, $\Omega_r$, $\Omega_\theta$ and $\Omega_\phi$, for general bound geodesic orbits until recently. Using the elegant Hamilton-Jacobi formalism, Schmidt~\cite{Schmidt2002Celestial} derived the fundamental frequencies without discussing the coupling of the $r$ and $\theta$ motions. Although his results show that we can expand an arbitrary function of the  particle's orbit  in a Fourier series, we can not estimate the Fourier components because of the coupling of the $r$ and $\theta$-motion. Mino~\cite{Mino:2003yg} showed that we can separate $r$ and $\theta$-motion if we use new time parameter $\lambda$ and derived the integral forms of the periods of both $r$ and $\theta$-motion with respect to $\lambda$, which is called Mino time. Combining Schmidt's method with Mino time, Drasco and Hughes~\cite{Drasco:2004} derived the fundamental frequencies and showed how the Fourier components of arbitrary functions of orbits with respect to Mino time can be computed because of the decoupling of both $r$ and $\theta$ motions. They also showed how from these results using Mino time, the Fourier components with respect to coordinate time can also be derived. Thanks to these results, one can compute gravitational waves from EMRIs. Despite a lot of works on geodesic motion, the analytical expressions of null or timelike geodesics in Kerr spacetime are still important subjects. Fast and accurate computation of timelike geodesics is required to study gravitational waves from EMRIs and construct efficient templates for LISA data analysis. Then, Fujita and Hikida~\cite{fujita2009efficient} derived analytical expressions for bound timelike geodesic orbits in Kerr spacetime, with Mino time as the independent variable. They also gave the analytical expressions of the fundamental frequencies in terms of the elliptic integrals. 

While the preceding analysis neglects the small body's mass (test-particle approximation), realistic extreme mass-ratio systems require treatment of gravitational perturbations to the Kerr background metric. The EOB formalism,
by including the mass-ratio corrections up to a certain order in the post-Newtonian (PN) expansion, can well describe the dynamical
evolution of binary black holes ~\cite{buonanno1999eff,buonanno2000transition}, and is widely used to construct the waveform templates for LIGO~\cite{taracchini2014effective,buonanno2007approaching,purrer2016frequency,husa2016frequency,khan2016frequency,chu2016accuracy,kumar2016accuracy,pan2014inspiral}. Note that the EOB formalism’s correction in the extreme-mass-ratio limit has not been guaranteed. However, as stated in~Ref.\cite{albanesi2021effective}, the extreme-mass-ratio limit plays a pivotal role in the EOB development, especially for what concerns waveforms and fluxes, which can be informed by and compared with numerical results ~\cite{Nagar:2006xv,Damour:2007xr,Bernuzzi:2010ty,Bernuzzi:2011bc,Bernuzzi:2010xj,Bernuzzi:2011aj,Harms:2014an,Nagar:2014kha,Harms:2015ixa,Harms:2016ctx,Lukes:2017st,Nagar:2019wrt}. In addition, the EOB formalism with the extreme-mass-ratio limit should be improved due to a lot of works dedicated to analytically calculating gravitational self-force terms and providing comparisons with numerical results ~\cite{Damour:2009sm,Barack:2010ny,Akcay:2012ea,Bini:2014ica,Bini:2015mza,Akcay:2015pjz,Barack:2019agd}. In the previous works\cite{zc2022}, we derived the analytically geometric solutions of the effective-one-body (EOB) dynamics with extreme-mass ratio limit. However, the r- and $\theta$- components of the geodesic equations remain coupled.

In this work, we derive analytical expressions for bound timelike geodesics in terms of Legendre elliptic integrals through appropriate coordinate transformations of $(r,\theta)$ and controlled approximations. This is the first time that the semi-analytical expressions of fundamental frequencies with mass-ratio correction are derived. With these analytical expressions at hand, we can perform Fourier expansions in Mino time $\lambda$  for any function expressed in terms of the coordinates $(r,\theta,\phi)$. These analytical expressions of bound timelike geodesic orbits with respect to Mino time should be helpful to calculate gravitational waves from EMRIs with mass-ratio correction by the Teukolsky equation. 

The organization of this paper is as follows. The basic knowledge of EOB formalism and deformed Kerr metric is introduced in the following section. In Sec.~\ref{sec:FF}, we present the derivation of analytical expressions for the fundamental frequencies governing bound geodesic orbits in Mino time. Since Mino time decouples the $r$ and $\theta$-motion, we also gave explicit expressions for three orbital frequencies $\Omega_r, ~\Omega_\theta, ~\Omega_\phi$ using the Fourier series expansion. Then, we checked the consistency of the analytical expressions comparing them with the numerical integration method. Finally, the last section contains our conclusions and discussions.

Throughout this paper we use geometric units $G = c = 1$, the units of time and length are the mass of
system $M$, and the units of linear and angular momentum are $\mu$ and $\mu M$, respectively, where $\mu$ is the reduced mass of the effective body.
\section{Effective-one-body Hamiltonian}\label{sec:geodesic-motion-kerr}

The Effective-One-Body (EOB) approach~\cite{buonanno1999eff,buonanno2000transition} to the general relativistic two-body problem maps the binary dynamics into an effective one-body system: a test particle moving in an external effective metric. This metric represents a deformation of the Schwarzschild (or Kerr) metric, parameterized by a deformation parameter that depends on the mass ratio of the original binary system. The
model synthesizes strong-field effects from the test-particle limit with finite mass-ratio corrections derived from the
PN approximation, while retaining the flexibility to incorporate nonperturbative insights from numerical relativity simulations~\cite{Bohe:2016gbl,taracchini2014effective,Damour:2001tu,Damour:2014sva,barausse2010improved}. By including mass-ratio corrections and spin effects, the EOB formalism allows for a more accurate description of the binary's dynamics, especially in the late stages of the inspiral, merger, and ringdown phases. We start by considering an EMRI system with a central Kerr black hole $m_1$ and inspiraling object $m_2$ (we assume
that it is nonspinning for simplicity, $m_2 \ll m_1$). For the moment, we neglect the radiation-reaction effects and focus on purely geodesic motion. The conservative orbital dynamics is derived via Hamilton's equations using the EOB Hamiltonian 
 \be
 H_{\rm EOB}=M\sqrt{1+2\nu (\hat{H}_{\rm eff}-1)}\,,\label{eq:HEOB}
 \ee
where $M=m_1+m_2$, $\nu=m_1m_2/M^2$, $\mu=\nu M$ and the reduced effective Hamiltonian $\hat{H}_{\rm eff}=H_{\rm eff}/\mu$. The effective metric, with $\nu$ as a deformation parameter, is written in the form
\begin{eqnarray}
\label{eq:EOBmetric}
g_{\rm eff}^{\alpha\beta}  \, P_{\alpha} \, P_{\beta} && \, = \frac{1}{r^2 + a^2 \cos^2 
\theta} \nonumber \\
\times&&\left[ \Delta_r (r) \, P_r^2 + P_{\theta}^2 + \frac{1}{\sin^2 \theta} \, 
(P_{\phi} 
+ a \sin^2 \theta \, P_t)^2 - \frac{1}{\Delta_t (r)} \, ((r^2 + a^2) \, P_t + a \, 
P_{\phi})^2 \right] \, ,
\end{eqnarray}
with
\be
\Delta_r (r) = \frac{r^2 \, A (u) + a^2}{D (u)} \ , \quad 
\Delta_t (r) = r^2 \, A(u) + a^2 \, .
\ee
where $a=|\bm{S}_{\rm{Kerr}}|/M$ is the effective Kerr parameter and $u = M/r$. The metric potentials $A$ and $D$ for the EOB model through 3PN 
order (for nonspinning metric) are given by~\cite{buonanno2000transition,damour2000determination}
\bes
\bea
A(u)&=&1-2u+2\nu u^3+\left(\frac{94}{3}-\frac{41 \pi ^2}{32}\right) \nu u^4\,,\\
D^{-1}(u)&=&1+6\nu u^2+2 \nu u^3(26-3\nu).
\eea
\ees
We use the formulation truncated at 3PN, since its deviation from the more accurate 5PN log-resummed potentials~\cite{steinhoff2016Apotential} remains below $0.01\nu$ within typical parameter regions, while its compact form is crucial for subsequent analytical calculations of the dynamics. To further refine the model, Barausse and Buonanno constructed an appropriate deformed Kerr metric such that the corresponding Hamiltonian exactly reproduces that of a spinning test particle in Kerr spacetime~\cite{barausse2010improved}. The deformed Kerr metric is given by
\bes
\bea
\label{def_metric_in}
g^{tt} &=& -\frac{\Lambda_t}{\Delta_t\,\Sigma}\,,\\
g^{rr} &=& \frac{\Delta_r}{\Sigma}\,,\\
g^{\theta\theta} &=& \frac{1}{\Sigma}\,,\\
g^{\phi\phi} &=& \frac{1}{\Lambda_t}
\left(-\frac{\widetilde{\omega}_{\rm fd}^2}{\Delta_t\,\Sigma}+\frac{\Sigma}{\sin^2\theta}\right)\,,\label{eq:gff}\\
g^{t\phi}&=&-\frac{\widetilde{\omega}_{\rm fd}}{\Delta_t\,\Sigma}\,.\label{def_metric_fin}
\eea
\ees
The quantities $\Sigma$, $\Delta_t$, $\Delta_r$, $\Lambda_t$, and $\widetilde{\omega}_{\rm fd}$
in Eqs.~(\ref{def_metric_in})-(\ref{def_metric_fin}) are given by 
\bes
\bea
\Sigma &=& r^2+a^2 \cos ^2\theta \,, \\
\label{deltat}
\Delta_t &=& r^2 A(u) + a^2\,, \\
\label{deltar}
\Delta_r &=& \Delta_t\, D^{-1}(u)\,,\\
\Lambda_t &=& \left(r^2+a^2\right)^2-a^2\Delta _{t}\sin ^2\theta \,,\\
\widetilde{\omega}_{\rm fd} &=& 2 a\, M\, r+ \omega_1^{\rm fd}\,\nu\,\frac{a M^3}{r}+ \omega_2^{\rm fd}\,\nu\,\frac{M a^3}{r}
\label{eq:omegaTilde}\,,
\eea
\ees
The parameters $\omega_1^{\rm fd}$ and $\omega_2^{\rm fd}$are adjustable parameters regulating the frame-dragging strength, with values approximately $-10$ and $20$, respectively, based on a preliminary EOB–numerical relativity comparison~\cite{rezzolla2008final,barausse2009predicting}. In addition to the metric, the effective action, $S_{\rm eff} = -\int \mu\, ds_{\rm eff}$, is introduced to govern the system’s dynamics, enabling a unified treatment of both the gravitational interaction and the emitted radiation. The effective one-body dynamics was given by an Hamilton-Jacobi equation of the form
\be
0 = \mu^2 + g_{\rm eff}^{\alpha \beta} \, P_{\alpha}^{\rm eff} \, P_{\beta}^{\rm eff}  \, ,
\ee
The effective Hamiltonian $H_{\rm eff}$ for a non-spinning particle in the deformed-Kerr metric then takes the form\cite{zc2022}
\begin{equation}
{H}_{\rm eff} = \frac{g^{t \phi}}{g^{tt}} P_\phi + \frac{1}{\sqrt{-g^{tt}}} \sqrt{\mu^2 + \Bigg[g^{\phi\phi}-\frac{(g^{t \phi})^2}{g^{tt}} \Bigg]P_\phi^2+g^{rr}P_r^2+g^{\theta\theta }P^2_\theta }\,,
\label{eq:Hnsdef}
\end{equation} 
the metric components have to be replaced with those of the deformed-Kerr metric (\ref{def_metric_in})--(\ref{def_metric_fin}).

\section{The fundamental frequencies}
\label{sec:FF}
We now present the derivation of analytical expressions for the fundamental frequencies governing bound geodesic orbits in Mino time. Section \ref{sec:mino} introduces the framework of deformed Kerr geodesics expressed in Mino time, while Sections \ref{sec:omega_r_theta} and \ref{sec:omega_t_phi} detail the systematic derivation of the radial/polar and temporal/azimuthal frequency components, respectively. The validity of these analytical expressions is verified in Section \ref{sec:check_omega} through comprehensive comparison with established numerical results in the literature.
\subsection{Geodesics in Mino time}
\label{sec:mino}
The geodesic equations for $r$, $\theta$, and $\phi$ are given by \cite{zc2022}
\bes\label{eomsEOBrphiEL}\begin{align}
\frac{dr}{dt}&=\frac{\partial E}{\partial P_{r}} =-\frac{g^{rr} \hat{P_r}}{E/M \left(g^{tt}\hat{H}_{\rm eff}-g^{t \phi}\hat{L}_z\right)}\,,  \label{eq:rdot}\\
\frac{d\theta}{dt}&=\frac{\partial E}{\partial P_{\theta}} =-\frac{g^{\theta \theta} \hat{P_{\theta}}}{E/M \left(g^{tt}\hat{H}_{\rm eff}-g^{t \phi}\hat{L}_z\right)}\,,\label{eq:thetadot}\\
\frac{d\phi}{dt}&=\frac{\partial E}{\partial P_{\phi}} =\frac{g^{t \phi}-\Big[g^{tt}g^{\phi\phi}-(g^{t \phi})^2 \Big]\frac{\hat{L}_z}{g^{tt} \hat{H}_{\rm eff}-g^{t \phi}\hat{L}_z}}{E/M g^{tt} }\,,\label{eq:phidot}
\end{align}
\ees
where the energy of the system is given by
\be
E=H_{\rm EOB},
\ee
and where $P_r$, $ P_{\phi}$, and $P_{\theta}$ are the canonical radial, azimuthal, and polar components of the angular momentum in the EOB gauge. We have defined the reduced momenta $\hat{P}=P/\mu$ and two constants of motion $\hat{H}_{\rm eff}=-P_t/\mu, \hat{L}_{z}={P}_{\phi }/\mu$(have been given in Eqs.(22a)-(22b) of \cite{zc2022}). 
Furthermore, $\hat{P_{\theta}},\hat{P_r}$ have been analytically obtained in \eqref{eq:ptheta}-\eqref{eq:pr}
\begin{widetext}
\bes
\bea
&\hat{P_{\theta}}^2=\hat{Q}-\cos ^2\theta  \bigg[a^2 \left(1-\hat{H}_{\rm eff}^2\right)+\cfrac{\hat{L}_{z}^2}{\sin^2 \theta}\bigg]\,\label{eq:ptheta},\\
&\hat{P_r}^2=\cfrac{\Big[a\hat{L}_{z}\!-\!(r^2+a^2)\hat{H}_{\rm eff}\Big]^2\!-\!\left( r^2 A(u)+a^2\right)\Big[r^2+(a\hat{H}_{\rm eff}-\hat{L}_{z})^2+\hat{Q}+2\cfrac{\widetilde{\omega}_{\rm fd}+a r^2\left(A(u)-1\right)}{r^2A(u)+a^2}\hat{H}_{\rm eff}\hat{L}_{z}-G(r)\hat{L}_{z}^2\Big]}{\left( r^2 A(u)+a^2\right)^2 D^{-1}(u) }\,\label{eq:pr}.
\eea
\ees
\end{widetext}
where
\be
G(r)=\frac{{\widetilde{\omega}_{\rm fd}}^2-a^2r^4\left(A(u)-1\right)^2}{\left(r^2A(u)+a^2\right)\left(r^2+a^2\right)^2} \,, 
\ee
and $\hat{Q}$ is the semi-Carter constant which has the same form as in the test-particle case\cite{zc2022}
\bea
\hat{Q} = \cos ^2\theta_{\rm min}  \bigg[a^2 \left(1-\hat{H}_{\rm eff}^2\right)+\cfrac{\hat{L_{z}}^2}{\sin^2 \theta_{\rm min}}\bigg]\,.\label{eq:carter}
\eea

We use Mino time as our time parameter describing these orbits. An interval of Mino time $d \lambda$ is related to an interval of coordinate time $dt$ by $d\lambda=\frac{g^{\theta \theta} }{ E/M (g^{t \phi}\hat{L}_z-g^{tt}\hat{H}_{\rm eff})}dt$, the geodesic equations become
\bes\label{mino}\begin{align}
\left(\frac{dr}{d\lambda}\right)^2&=\Delta_r^2 \hat{P}^2_r=\left(1+6\nu u^2+2 \nu u^3(26-3\nu)\right)R(r)\,,  \label{eq:rmino}\\
\left(\frac{d\cos\theta}{d\lambda}\right)^2&=\hat{P_{\theta}}^2=\Theta(\cos\theta)\,\label{eq:thetamino}\,, \\
\frac{d\phi}{d\lambda}&=\frac{\widetilde{\omega}_{\rm fd}\hat{H}_{\rm eff}-a^2\hat{L}_{z}}{\left(r^2A(u)+a^2\right)} +\frac{\hat{L}_{z}}{\sin^2\theta}-\frac{4a^2\nu\left( 20a^2-8+(\frac{94}{3}-\frac{41 \pi ^2}{32})u\right)\hat{L}_{z}}{\left(r^2A(u)+a^2\right)\left(r^2+a^2\right)^2}\,,\label{eq:phimino}\\
\frac{dt}{d\lambda}&=(E/M)\big[\frac{(r^2+a^2)^2\hat{H}_{\rm eff}-\widetilde{\omega}_{\rm fd}\hat{L}_{z}}{\left(r^2A(u)+a^2\right)}-a^2\hat{H}_{\rm eff}\sin^2\theta\big]\,\label{eq:tmino},
\end{align}
\ees
where the denominator's third term in Eq.\eqref{eq:phimino} uses an approximation (originally $\Delta_t\Lambda_t$) which \cite{zc2022} confirms to be well within safe limits, and where
\bes
\begin{align}
\label{completeR}
R(r)&=\Big[a\hat{L}_{z}\!-\!(r^2\!+\!a^2)\hat{H}_{\rm eff}\Big]^2\!-\!\left( r^2 A(u)\!+\!a^2\right)\Big[r^2\!+\!(a\hat{H}_{\rm eff}\!-\!\hat{L}_{z})^2\!+\!\hat{Q}\!+\!2\cfrac{\widetilde{\omega}_{\rm fd}\!+\!a r^2\left(A(u)\!-\!1\right)}{r^2A(u)+a^2}\hat{H}_{\rm eff}\hat{L}_{z}\!-\!G(r)\hat{L}_{z}^2\Big], \\
\Theta(\cos\theta)&=\beta^2(z_{-}-\cos^2\theta)(z_{+}-\cos^2\theta).
\end{align}
\ees
and where $\beta^2=a^2(1-\hat{H}_{\rm eff}^2),$
$z_{-}=\cos^2\theta_{\T{min}},$ $z_{+}=\hat{Q}/(\beta^2 z_{-})$. For the bound orbits, $r(\lambda)$ and $\theta(\lambda)$ become periodic functions that are independent of each other. We note that the above equations of motion extend the standard test-particle geodesic formulation\cite{Mino:2003yg} through the inclusion of mass-ratio corrections. The fundamental periods for the radial and polar motion, $\Lambda_{r}$ and $\Lambda_{\theta}$, are defined as(in the following calculation, we neglect the $\mathcal{O}(\nu^2)$ term)
\bes
\begin{align}
\label{periodr}
\Lambda_r&=2\int_{\rm r_{\rm min}}^{\rm r_{\rm max}}\frac{\rm dr}{\sqrt{(1+6\nu u^2+2 \nu u^3(26-3\nu))R(r)}}=2\int_{\rm r_{\rm min}}^{\rm r_{\rm max}}\left(\frac{1}{\sqrt{R(r)}}-\frac{3\nu }{r^2\sqrt{R(r)}}-\frac{26\nu }{r^3\sqrt{R(r)}}\right)\rm dr,\\
\Lambda_\theta&=4\int_{0}^{\cos\theta_{\rm min}}
\frac{\rm d\cos\theta}{\sqrt{\Theta(\cos\theta)}}.
\label{periods} 
\end{align}
\ees

The angular frequencies of the radial and polar motion then become
\begin{eqnarray}
\Upsilon_r=\frac{2\pi}{\Lambda_r}\,,\qquad
\Upsilon_\theta=\frac{2\pi}{\Lambda_\theta}.
\label{Upsilon}
\end{eqnarray}
We define the angle variables as $w_r=\Upsilon_r\lambda$ and $w_\theta=\Upsilon_\theta\lambda$. The $r$-dependent and $\theta$-dependent functions are $2\pi$-periodic in $w_r$ and $w_\theta$, respectively.

From the geodesic equations Eqs.\eqref{eq:phimino} and \eqref{eq:tmino}, $\rm dt/\rm d \lambda$ and $\rm d\phi/\rm d \lambda$ are functions only of $r$ and $\theta$. This means that they can be expanded in a Fourier series:
\begin{eqnarray}
\label{geodesic_phi_fourier}
\frac{dt}{d\lambda} &=&\sum_{k,n}T_{k,n}
e^{-ik\Upsilon_r\lambda}e^{-in\Upsilon_\theta\lambda}, \\
\label{geodesic_t_fourier}
\frac{d\phi}{d\lambda} &=&\sum_{k,n}\Phi_{k,n}
e^{-ik\Upsilon_r\lambda}e^{-in\Upsilon_\theta\lambda},
\end{eqnarray}
where
\begin{eqnarray}
T_{k,n}&=&\frac{\sqrt{1+2\nu(\hat{H}_{\rm eff}-1)}}{2\pi}\left(\int_0^{2\pi}\frac{(r^2+a^2)^2\hat{H}_{\rm eff}-\widetilde{\omega}_{\rm fd}\hat{L}_{z}}{\Delta_t }e^{i k w_r}dw_r +\int_0^{2\pi}a^2\hat{H}_{\rm eff}(\cos^2\theta-1 )e^{i n w_\theta}dw_\theta\right)
, \label{Tkn}\\
\Phi_{k,n}&=&\frac{1}{2\pi}\left(\int_0^{2\pi}\frac{\widetilde{\omega}_{\rm fd}\hat{H}_{\rm eff}-a^2\hat{L}_{z}}{\left(r^2A(u)+a^2\right)} -\frac{4a^2\nu\left( 20a^2-8+(\frac{94}{3}-\frac{41 \pi ^2}{32})u\right)\hat{L}_{z}}{\left(r^2A(u)+a^2\right)\left(r^2+a^2\right)^2}e^{i k w_r}dw_r +\int_0^{2\pi}\frac{\hat{L}_{z}}{1-\cos^2\theta}e^{i n w_\theta}dw_\theta\right). \label{Phikn}
\end{eqnarray}
Since $T_{k,n}=0$ and $\Phi_{k,n}=0$ in the case of $k\neq 0$ and $n\neq 0$, we only need to compute $T_{k,0}, T_{0,n}, \Phi_{k,0}$ and $\Phi_{0,n}$,
which can be easily obtained via one-dimensional Fourier transformations. It is convenient to rewrite \eqref{geodesic_phi_fourier}\eqref{geodesic_t_fourier}
\begin{eqnarray}
\frac{dt}{d\lambda} &=&
\Gamma+\sum_{k\neq 0}T_{k,0}e^{-ikw_r}+\sum_{n\neq 0}T_{0,n}e^{-inw_\theta}, \\
\Gamma&\equiv& T_{00},\label{T00}\\
\frac{d\phi}{d\lambda} &=&
\Upsilon_\phi+\sum_{k\neq 0}\Phi_{k,0}e^{-ikw_r}+\sum_{n\neq 0}\Phi_{0,n}e^{-inw_\theta},\\
\Upsilon_\phi&\equiv& \Phi_{00}\label{Phi00}.
\end{eqnarray}
Using these results, it is simple to integrate for $t(\lambda)$ and $\phi(\lambda)$:
\begin{eqnarray}
t(\lambda)&=&\Gamma\lambda
+\sum_{k\neq 0}\frac{iT_{k,0}}{k\Upsilon_r}e^{-ikw_r}
+\sum_{n\neq 0}\frac{iT_{0,n}}{n\Upsilon_\theta}e^{-inw_\theta}, 
\label{geodesic_t_fourier2} \\
\phi(\lambda)&=&\Upsilon_\phi\lambda
+\sum_{k\neq 0}\frac{i\Phi_{k,0}}{k\Upsilon_r}e^{-ikw_r}
+\sum_{n\neq 0}\frac{i\Phi_{0,n}}{n\Upsilon_\theta}e^{-inw_\theta}.
\label{geodesic_phi_fourier2} 
\end{eqnarray}
The two variables, $\Gamma$ and $\Upsilon_\phi$, represent the average rates of change of $t$ and $\phi$ as functions of $\lambda$, respectively. We note that the frequencies with respect to $\lambda$ are related to 
the frequencies with distant observer time as~\cite{Drasco:2004}
\begin{eqnarray}
\Omega_{r}=\frac{\Upsilon_{r}}{\Gamma},\qquad
\Omega_{\theta}=\frac{\Upsilon_{\theta}}{\Gamma},\qquad
\Omega_{\phi}=\frac{\Upsilon_{\phi}}{\Gamma}.
\label{eq:Omega_r_th_phi}
\end{eqnarray}
In the following subsections, we discuss the analytical expressions for these frequencies. 

\subsection{Frequencies of $r$ and $\theta$-motion}
\label{sec:omega_r_theta}
In this subsection, we derive the analytical expressions for 
the frequencies of $r$ and $\theta$-motion, $\Upsilon_r$ and $\Upsilon_\theta$, 
using Eqs.\eqref{periodr}\eqref{periods}.
First, to solve the differential equations for $r$ and $\cos\theta$, we need to convert $R(r)$ into a quartic polynomial like $\Theta(\cos\theta)$. We approximate $R(r)$ by replacing it with the factored form $\mathcal{R}(r)$, thus obtaining a quartic polynomial expression
\begin{align*}
\label{factor}
\mathcal{R}(r)&=\Big[a\hat{L}_{z}-(r^2+a^2)\hat{H}_{\rm eff}\Big]^2-\left( r^2-2M r+a^2\right)\Big[r^2+(a\hat{H}_{\rm eff}-\hat{L}_{z})^2+\hat{Q}\Big]-C_\mathcal{R} \nu\\
&=(1 - \hat{H}_{\rm eff}^2)(r_1 - r)(r - r_2)(r - r_3)(r - r_4)\sim R(r),
\end{align*}
where 
\begin{align}
 r_{1} = \frac{p}{1-{e}},\quad
 r_{2} &= \frac{p}{1+{e}},\quad
 r_{3} = \frac{(A+B)+\sqrt{(A+B)^2-4AB}}{2},\quad
 r_{4} = \frac{AB}{r_3},\cr
 A+B &= \frac{2}{1-{\hat{H}_{\rm eff}}^2} - (r_1+r_2),\quad
 AB  = \frac{a^2\hat{Q}+C_\mathcal{R} \nu}{(1-{\hat{H}_{\rm eff}}^2)\,r_1r_2}.
\end{align}
We note that 
two zero points,
$r_{1}(r_{\rm max})$ and $r_{2}(r_{\rm min})$, of $R(r)$ are apoapsis and periapis respectively and $C_\mathcal{R}$ is an adjustable parameter which calculate by

\begin{align}
 \int_{\rm r_2}^{\rm r_1}\frac{dr}{\sqrt{\mathcal{R} (r)}}=\int_{\rm r_2}^{\rm r_1}\frac{dr}{\sqrt{R(r)}}
\end{align}

\begin{figure}[!h]
\begin{center}

\includegraphics[width=0.8\columnwidth]{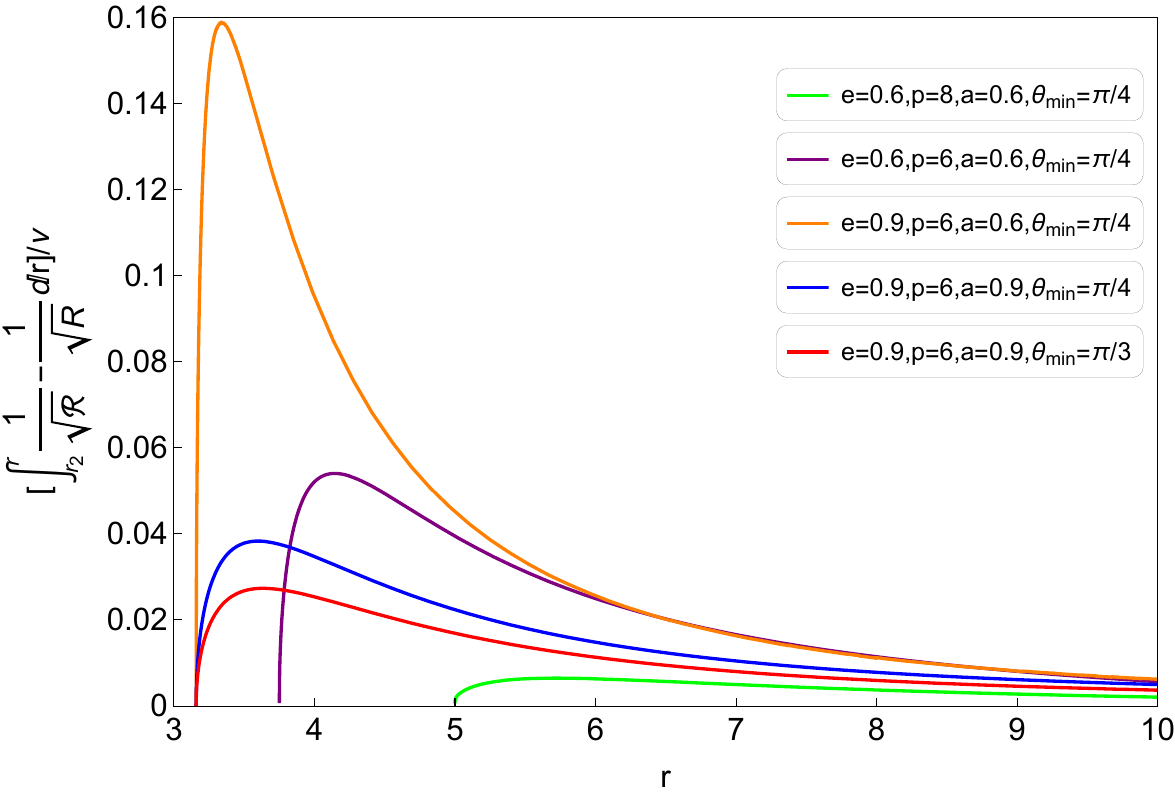}
\caption{\emph{Fitting error of the integral $\int_{\rm r_2}^{\rm r}\frac{dr}{\sqrt{R(r)}}$.} Replacing ${R(r)}$ with $\mathcal{R}(r)$ provides an highly precise approximation. Even if we choose extreme orbital parameters that bring the periapsis $r_2$ near LSO(orange line), the error is less than $0.16\nu$.}
\label{fig:1}
\end{center}
\end{figure}

For example, for this set of orbital parameters $(e=0.6,p=8,a=0.6,\theta_{min}=\pi/4)$, $C_\mathcal{R}=57$(this value remains nearly the same across different $\nu$). As shown in Fig.~\ref{fig:1}, the integral error between $1/R(r)$ and $1/\mathcal{R}(r)$ depends mainly on the value of $r$ and is maximal at a value of $r$ close to $r_2$.  As $r$ increases, this error asymptotically approaches zero. One can see that even when selecting extreme orbital parameters that bring the periapsis close to the LSO(Last Stable Orbit, for the orange line, $r_{\rm{LSO}}=2.9M$), the integral error is below $0.16\nu$. For most of the orbital parameter space, this approximation introduces errors one order smaller than the mass ratio. We believe that replacing $R(r)$ with $\mathcal{R}(r)$ retains enough precision for EMRIs. 

Since both $\mathcal{R}(r)$ and $\Theta(\cos\theta)$ are fourth-order polynomials,
Eqs.\eqref{periodr}\eqref{periods} can be expressed in terms of the
elliptic integrals, respectively
\begin{align}
 \Lambda_{r} &=
 \frac{4}{\sqrt{(1-{\hat{H}_{\rm eff}}^2)(r_1-r_3)(r_2-r_4)}}\Big[ K(k_r)-3\nu\frac{Z_2|_{\varphi=\pi/2}}{ r_2^2}-26\nu\frac{Z_3|_{\varphi=\pi/2}}{ r_2^3}\Big],\quad\label{eq:period_r}\\
\Lambda_{\theta} &=
\frac{4 K(k_{\theta})}{\sqrt{\hat{Q}/ z_{-}}}.
\label{eq:period_theta}
\end{align}
where
\begin{align}
k_{r}&=\frac{r_1-r_2}{r_1-r_3}\frac{r_3-r_4}{r_2-r_4},\\
k_{\theta}&=\frac{z_{-}}{z_{+}},\\
\alpha&=\frac{r_1-r_2}{r_1-r_3},\\
Z_2&=\frac{\alpha^2F(\varphi,k_r)+2\alpha(r_3\alpha/r_2-\alpha)\Pi(\varphi,r_3\alpha/r_2,k_r)+(r_3\alpha/r_2-\alpha)^2V_2}{(r_3\alpha/r_2)^2},\label{Z2}\\
Z_3&=\frac{\alpha^3F(\varphi,k_r)+3\alpha^2(r_3\alpha/r_2-\alpha)\Pi(\varphi,r_3\alpha/r_2,k_r)+3\alpha(r_3\alpha/r_2-\alpha)^2V_2+(r_3\alpha/r_2-\alpha)^3V_3}{(r_3\alpha/r_2)^3},\label{Z3}\\
V_2&=\frac{1}{2(\frac{r_3}{r_2}\alpha\!-\!1)(k_r\!-\!\frac{r_3}{r_2}\alpha)}\Bigg[\frac{r_3}{r_2}\alpha E(\varphi,k_r)+(k_r\!-\!\frac{r_3}{r_2}\alpha)F(\varphi,k_r)\!+\!(2\frac{r_3}{r_2}\alpha k_r\!+\!2\frac{r_3}{r_2}\alpha\!-\!\frac{r_3^2}{r_2^2}\alpha^2\!-\!3k_r)\Pi(\varphi,r_3\alpha/r_2,k_r)\nonumber\\
&-\frac{\frac{r_3^2}{r_2^2}\alpha^2\T{sn}(\varphi;k_r)\T{cn}(\varphi;k_r)
 \T{dn}(\varphi;k_r)}{1\!-\!\frac{r_3}{r_2}\alpha \T{sn}^2(\varphi;k_r)}\Bigg],\\
V_3&=\frac{1}{4(1\!-\!\frac{r_3}{r_2}\alpha)(k_r\!-\!\frac{r_3}{r_2}\alpha)}\Bigg[k_r F(\varphi,k_r)\!+\!2(\frac{r_3}{r_2}\alpha k_r\!+\!\frac{r_3}{r_2}\alpha\!-\!3k_r)\Pi(\varphi,r_3\alpha/r_2,k_r)\!+\!3(\frac{r_3^2}{r_2^2}\alpha^2-2\frac{r_3}{r_2}\alpha k_r\!-\!2\frac{r_3}{r_2}\alpha\!+\!3k_r)V_2\nonumber\\
&+\frac{\frac{r_3^2}{r_2^2}\alpha^2\T{sn}(\varphi;k_r)\T{cn}(\varphi;k_r)
 \T{dn}(\varphi;k_r)}{\left(1\!-\!\frac{r_3}{r_2}\alpha \T{sn}^2(\varphi;k_r)\right)^2}\Bigg],
\end{align}
the function $\T{sn}(\varphi,k)$ 
, $\T{cn}(\varphi,k)$ and  $\T{dn}(\varphi,k)$ are Jacobi elliptic functions and $K(k)$, $E(k)$, $\Pi(\varphi,k)$ are, respectively, the complete elliptical
integrals of the first, second and third kind,
\bes
\bea
  \label{eq:ell_integr_k}
  K(k) & =&\int_{0}^{\pi/2}\!\frac{\rmd \chi}{\sqrt{1-k\sin^{2}\chi}}, \\
  \label{eq:ell_integr_e}
  E(k) & =&\int_{0}^{\pi/2}\!\sqrt{1-k\sin^{2}\chi}\,\rmd \chi, \\
  \label{eq:ell_integr_pi}
  \Pi(\varphi,k) & =&
  \int_{0}^{\pi/2}\!\frac{\rmd\chi}{\left(1-\varphi\sin^{2}\chi\right)
  \sqrt{1-k\sin^{2}\chi}}.
\eea
\ees
The orbital frequencies of radial and polar motion with respect to $\lambda$ are calculated using \eqref{Upsilon}
\begin{align}
\Upsilon_r &=
 \frac{\pi\sqrt{(1-{\hat{H}_{\rm eff}}^2)(r_1-r_3)(r_2-r_4)}}{2\Big[ K(k_r)-3\nu\frac{Z_2|_{\varphi=\pi/2}}{ r_2^2}-26\nu\frac{Z_3|_{\varphi=\pi/2}}{ r_2^3}\Big]},\label{eq:fre_r}\\
\Upsilon_\theta &=
\frac{\pi \sqrt{\hat{Q}/ z_{-}}}{2K(k_{\theta})}.
\label{eq:fre_theta}
\end{align}
We note that the above equations have similar forms as the test particle ones given in Appendix of \cite{fujita2009efficient}. However, the mass-ratio corrections have to be encoded in each variable. 
\subsection{Frequencies of $t$ and $\phi$-motion}
\label{sec:omega_t_phi}
In this subsection, we derive the explicit expressions for 
the frequencies of $t$ and $\phi$-motion, $\Gamma$ and $\Upsilon_\phi$, 
using the Fourier series expansion Eqs.\eqref{Tkn} and \eqref{Phikn}. Essentially, we seek to express the coordinates $r$ and $\theta$ as functions of their conjugate angle variables $w_r$ and $w_\theta$ respectively. Let the solutions of Eqs.(\ref{eq:rmino}) and (\ref{eq:phimino}) in terms of 
$r$ or $\theta$ be $\lambda(r)$ and $\lambda(\theta)$,
respectively. We obtain $\lambda(r)$ and $\lambda(\theta)$ as
\begin{align}
\lambda(r)&= \int_{}^{\rm r}\left(\frac{1}{\sqrt{\mathcal{R}(r')}}-\frac{3\nu }{r^2\sqrt{\mathcal{R}(r')}}-\frac{26\nu }{r^3\sqrt{\mathcal{R}(r')}}\right)\rm dr',\\
\lambda(\theta)&=
\int^{\cos\theta}\frac{\rmd\cos\theta '}{\sqrt{\Theta(\cos\theta ')}}.
\label{eq:lam_t_phi_r}
\end{align}
Since the period of $r$ and $\theta$-motion with respect to $\lambda$ is 
$\Lambda_{r,\theta}=2\pi/\Upsilon_{r,\theta}$, 
we map $\lambda$ to $\lambda^{({r,\theta})}$ as 
$\lambda^{({r,\theta})}=\lambda-2\pi[\Upsilon_{r,\theta}\lambda/2\pi]/\Upsilon_{r,\theta}$, 
where $[\cdots]$ is the floor function, in the following subsections. We set the initial
value of $r(\lambda)$ and $\theta(\lambda)$ satisfies $r(\lambda=0)=r_2$, $\cos\theta(\lambda=0)=\pi/2$, respectively.
Then the functions $\lambda^{(r)}(r)$ and $\lambda^{(\theta)}(\theta)$ can be
expressed as 
\begin{align}
  \lambda^{(r)}(r) &=
  \left\{\begin{array}{ll} 
  \lambda^{(r)}_0(r)    \,\,\,\,\, &r:r_2\rightarrow r_1,\\
  2\lambda^{(r)}_0(r_1)-\lambda^{(r)}_0(r)    \,\,\,\,\, &r:r_1\rightarrow r_2,
  \end{array}\right.\cr
  \lambda^{(\theta)}(\theta) &=
  \left\{\begin{array}{ll} 
  \lambda^{(\theta)}_0(\theta)    \,\,\,\,\, &\theta:\frac{\pi}{2}\rightarrow \theta_{\rm min},\\
  2\lambda^{(\theta)}_0(\theta_{\rm min})-\lambda^{(\theta)}_0(\theta)    \,\,\,\,\, &\theta:\theta_{\rm min}\rightarrow \pi-\theta_{\rm min},\\
  4\lambda^{(\theta)}_0(\theta_{\rm min})+\lambda^{(\theta)}_0(\theta)    \,\,\,\,\, &\theta:\pi-\theta_{\rm min}\rightarrow \frac{\pi}{2},
  \end{array}\right.
\label{eq:sol_eom1}
\end{align}
where
\begin{align}
 \lambda^{(r)}_0(r) &= \frac{2}{\sqrt{(1-\hat{H}_{\rm eff}^2)(r_1-r_3)(r_2-r_4)}}
\bigg[F\left(\arcsin y_r
,k_r\right)-3\nu\frac{Z_2}{ r_2^2}-26\nu\frac{Z_3}{ r_2^3}\bigg],\label{lambda0r}\\
 \lambda^{(\theta)}_0(\theta) &= \frac{1}{\sqrt{\hat{Q}/ z_{-}}}
 F\left(\arcsin y_\theta,k_\theta\right),
\end{align}
here 
\begin{align}
 y_{r}&=\frac{r_1-r_3}{r-r_2}\frac{r_1-r_2}{r-r_3},\label{yr}\\
 y_{\theta}&=\frac{\cos{\theta}}{\sqrt{z_{-}}},
\end{align}
and $Z_2,Z_3$ are given in \eqref{Z2}\eqref{Z3}, where $\varphi=\arcsin y_r$. 
Furthermore, Eq.~\eqref{eq:sol_eom1} can be solved inversely. For the test-particle limit $\nu \to 0$, the expression of $\lambda^{(r)}_0(r)$ go back to the form in Kerr spacetime. The radial and polar motion can be expressed in terms of Jacobi elliptic functions\cite{fujita2009efficient}
\begin{align}
y_r'&=\text{sn}\left(\varphi_r(\omega_r),k_r\right),\\
y_\theta&=\text{sn}\left(\varphi_\theta(\omega_\theta),k_\theta\right)
\end{align}
where
\begin{align}
  \varphi_r(w_r)&=
  \left\{\begin{array}{ll} 
  w_r\frac{K(k_r)}{\pi}, \,\,\,\,\, &(0\le w_r \le \pi)\\
  (2\pi-w_r)\frac{K(k_r)}{\pi}, \,\,\,\,\, &(\pi\le w_r \le 2\pi)
  \end{array}\right.\cr
  \varphi_\theta(w_\theta)&=
  \left\{\begin{array}{ll} 
  w_\theta\frac{2K(k_\theta)}{\pi}, \,\,\,\,\, &(0\le w_\theta \le \frac{\pi}{2})\\
  (\pi-w_\theta)\frac{2K(k_\theta)}{\pi}, \,\,\,\,\, &(\frac{\pi}{2}\le w_\theta \le \frac{3\pi}{2})\\
  (w_\theta-2\pi)\frac{2K(k_\theta)}{\pi}. \,\,\,\,\, &(\frac{3\pi}{2}\le w_\theta \le \pi)
  \end{array}\right.
\end{align}
the $r$ and $\theta$, denoted by the angle variables $w_{r}$ and $w_{\theta}$ respectively, are given by
\begin{align}
r(w_r)'&=\frac{r_3(r_1-r_2)\,\text{sn}^2\left(\varphi_r(\omega_r),k_r\right)-r_2(r_1-r_3)}
 {(r_1-r_2)\,\text{sn}^2\left(\varphi_r(\omega_r),k_r\right)-(r_1-r_3)},\\
[\cos\theta](w_\theta)&= \sqrt{z_{-}}\,\T{sn}(\varphi_\theta(w_\theta);k_\theta).
\label{eq:thetaw}
\end{align}
In order to drive the expression of 
$r(w_r)$ in deformed-Kerr spacetime, we perform a second-order Taylor expansion of the equation $\Upsilon_r\lambda^{r}(r)-w_r=0$(equivalently $F\left(\arcsin y_r
,k_r\right)-3\nu\frac{Z_2}{ r_2^2}-26\nu\frac{Z_3}{ r_2^3}-\frac{w _r}{\pi }\Big[ K(k_r)-3\nu\frac{Z_2|_{\varphi=\pi/2}}{ r_2^2}-26\nu\frac{Z_3|_{\varphi=\pi/2}}{ r_2^3}\Big]=0$) about the Kerr solution $y_r'=\text{sn}\left(\varphi_r(\omega_r)k_r\right)$:
\be
F_1+F_2(y_r-x)+F_3(y_r-x)^2=0,\label{FF}
\ee
where
\bes
\bea
x&=&y_r'=\text{sn}\left(\varphi_r(\omega_r),k_r\right),\\
\mathcal{F}_1&=&F\left(\arcsin{x},k_r\right)-\frac{w_r}{\pi }\Big[ K(k_r)-3\nu\frac{Z_2|_{\varphi=\pi/2}}{ r_2^2}-26\nu\frac{Z_3|_{\varphi=\pi/2}}{ r_2^3}\Big]-\frac{3 \nu}{ r_3^2}  \Bigg[F\left(\arcsin{x},k_r\right) +2 (\frac{r_3}{r_2}-1) \Pi (\frac{\alpha  r_3}{r_2};\arcsin{x},k_r)\nonumber\\
&&+\frac{\left(\frac{r_3}{r_2}-1\right)^2}{2 \left(\frac{\alpha  r_3}{r_2}-1\right) \left(k_r-\frac{\alpha  r_3}{r_2}\right)} \bigg(\frac{\alpha  r_3 E\left(\arcsin{x},k_r\right)}{r_2}+(k_r-\frac{\alpha  r_3}{r_2}) F\left(\arcsin{x},k_r\right)-\frac{\alpha ^2 r_3^2 x \sqrt{1-x^2} \sqrt{1-x^2 k_r}}{r_2 \left(r_2-\alpha  r_3 x^2\right)}\nonumber\\
&&+(\frac{2 \alpha  r_3 (k_r+1)}{r_2}-3 k_r-\frac{\alpha ^2 r_3^2}{r_2^2}) \Pi (\frac{\alpha  r_3}{r_2};\arcsin{x},k_r)\bigg) \Bigg]-\frac{26 \nu}{ r_3^2}  \Bigg[F\left(\arcsin{x},k_r\right)+3 (\frac{ r_3}{r_2}-1 ) \Pi (\frac{\alpha  r_3}{r_2};\sin ^{-1}(x),k_r) \nonumber\\
&&+\frac{\left(\frac{r_3}{r_2}-1\right)^2}{4\left(\frac{\alpha  r_3}{r_2}-1\right) \left(k_r-\frac{\alpha  r_3}{r_2}\right)}\bigg[ 6\bigg(\frac{\alpha  r_3 E\left(\arcsin{x},k_r\right)}{r_2}+(k_r-\frac{\alpha  r_3}{r_2}) F\left(\arcsin{x},k_r\right)-\frac{\alpha ^2 r_3^2 x \sqrt{1-x^2} \sqrt{1-x^2 k_r}}{r_2 \left(r_2-\alpha  r_3 x^2\right)}\nonumber\\
&&+(\frac{2 \alpha  r_3 (k_r+1)}{r_2}-3 k_r-\frac{\alpha ^2 r_3^2}{r_2^2}) \Pi (\frac{\alpha  r_3}{r_2};\arcsin{x},k_r)\bigg)-(\frac{r_3}{r_2}-1)\bigg(k_r F\left(\arcsin{x},k_r\right)+\frac{\alpha ^2 r_3^2 x \sqrt{1-x^2} \sqrt{1-x^2 k_r}}{\left(r_2-\alpha  r_3 x^2\right)^2}\nonumber\\
&&+2(\frac{\alpha  r_3 (k_r+1)}{r_2}-3 k_r) \Pi (\frac{\alpha  r_3}{r_2};\arcsin{x},k_r)-\frac{3\left(\frac{2 \alpha  r_3 (k_r+1)}{r_2}-3 k_r-\frac{\alpha ^2 r_3^2}{r_2^2}\right) }{2 \left(\frac{\alpha  r_3}{r_2}-1\right) \left(k_r-\frac{\alpha  r_3}{r_2}\right)} \Big((k_r-\frac{\alpha  r_3}{r_2}) F\left(\arcsin{x},k_r\right)\nonumber\\
&&+\frac{\alpha  r_3 E\left(\arcsin{x},k_r\right)}{r_2}-\frac{\alpha ^2 r_3^2 x \sqrt{1-x^2} \sqrt{1-x^2 k_r}}{r_2 \left(r_2-\alpha  r_3 x^2\right)}+(\frac{2 \alpha  r_3 (k_r+1)}{r_2}-3 k_r-\frac{\alpha ^2 r_3^2}{r_2^2}) \Pi (\frac{\alpha  r_3}{r_2};\arcsin{x},k_r)\Big)\bigg)\bigg]\Bigg],\\
\mathcal{F}_2&=&\frac{1}{\sqrt{1-x^2} \sqrt{1-x^2 k_r}}-\frac{3 \nu  \left(\alpha ^2 x^4-2 \alpha  x^2+1\right)}{\sqrt{1-x^2} \sqrt{1-x^2 k_r} \left(\alpha  r_3 x^2-r_2\right)^2}-\frac{26 \nu  \left(\alpha ^3 x^6-3 \alpha ^2 x^4+3 \alpha  x^2-1\right)}{\sqrt{1-x^2} \sqrt{1-x^2 k_r} \left(\alpha  r_3 x^2-r_2\right){}^3},\\
\mathcal{F}_3&=&-\frac{x \left(2 x^2 k_r-k_r-1\right)}{2\left(1-x^2\right)^{3/2}  \left(1-x^2 k_r\right)^{3/2} }\nonumber\\
&&
+\frac{3 \nu  x \left(\alpha  x^2-1\right) }{2  \left(1-x^2\right)^{3/2}  \left(1-x^2 k_r\right)^{3/2} \left(\alpha  r_3 x^2-r_2\right)^3}\Big[r_2 \left(4 \alpha +k_r \left(x^2 \left(2 \alpha  x^2-3 \alpha+2\right)-1\right)-3 \alpha  x^2-1\right)\nonumber\\
&&
+\alpha  r_3 \left(x^2 k_r (2 \alpha  x^4-(\alpha +6) x^2+5)-\alpha  x^4+5 x^2-4\right)\Big]+\frac{13 \nu  x \left(\alpha  x^2-1\right)^2 }{\left(1-x^2\right)^{3/2}  \left(1-x^2 k_r\right)^{3/2} \left(\alpha  r_3 x^2-r_2\right)^4}\Big[r_2 (6 \alpha\nonumber\\
&& +k_r \left(x^2 \left(4 \alpha  x^2-5 \alpha+2\right)-1\right)-5 \alpha  x^2-1)+\alpha  r_3 \left(x^2 k_r \left(2 \alpha  x^4-(\alpha +8) x^2+7\right)-\alpha  x^4+7 x^2-6\right)\Big].
\eea
\ees
Solving \eqref{FF} and \eqref{yr}, we obtain $y_r$ and $r(w_r)$ as
\begin{align}
 y_r&=\frac{2 \mathcal{F}_3 x+\sqrt{\mathcal{F}_2^2-4 \mathcal{F}_1 \mathcal{F}_3}-\mathcal{F}_2}{2 \mathcal{F}_3},\\
 r(w_r)&=\frac{r_3(r_1-r_2)\,y_r^2-r_2(r_1-r_3)}
 {(r_1-r_2)\,y_r^2-(r_1-r_3)}.
\label{eq:rwr}
\end{align}
By substituting $r$ and $\cos\theta$ in Eqs.\eqref{Tkn} and \eqref{Phikn} with the expressions from Eqs.\eqref{eq:rwr} and \eqref{eq:thetaw}, we obtain the the frequencies of $t$ and $\phi$-motion, $\Gamma$ and $\Upsilon_\phi$
\begin{eqnarray}
\Gamma&=&\frac{\sqrt{1+2\nu(\hat{H}_{\rm eff}-1)}}{2\pi}\left(\int_0^{2\pi}\frac{(r^2+a^2)^2\hat{H}_{\rm eff}-\widetilde{\omega}_{\rm fd}\hat{L}_{z}}{\Delta_t }dw_r +\int_0^{2\pi}a^2\hat{H}_{\rm eff}(\cos^2\theta-1 )dw_\theta\right)
, \label{Gamma}\\
\Upsilon_\phi&=&\frac{1}{2\pi}\left(\int_0^{2\pi}\frac{\widetilde{\omega}_{\rm fd}\hat{H}_{\rm eff}-a^2\hat{L}_{z}}{\left(r^2A(u)+a^2\right)} -\frac{4a^2\nu\left( 20a^2-8+(\frac{94}{3}-\frac{41 \pi ^2}{32})u\right)\hat{L}_{z}}{\left(r^2A(u)+a^2\right)\left(r^2+a^2\right)^2}dw_r +\int_0^{2\pi}\frac{\hat{L}_{z}}{1-\cos^2\theta}dw_\theta\right). \label{Upsilonphi}
\end{eqnarray}
Combining \eqref{eq:Omega_r_th_phi},  \eqref{eq:fre_r},\eqref{eq:fre_theta}, \eqref{Gamma} and \eqref{Upsilonphi}, we can derive the orbital frequencies with respect to observer time, 
$\Omega_r$, $\Omega_\theta$ and $\Omega_\phi$. 

A broad class of functions characterizing black hole orbital dynamics can be fully parametrized by the fundamental frequencies $\Omega_r$ and $\Omega_\theta$. Any function of the form $f[r(t),\theta(t)]$ (a common functional form for black hole orbits, since the metric is
independent of both $t$ and $\phi$) can be expanded as\cite{Drasco:2004}
\begin{equation}
f[r(t),\theta(t)] = \sum_{kn} f_{kn}
e^{-ik\Omega_\theta t}e^{-in\Omega_r t}\;.
\label{eq:f_expand_t}
\end{equation}
The Fourier expansion coefficients ${\tilde f}_{kn}$ of any function of the form $f[r(\lambda),\theta(\lambda)]$ is given by
\begin{equation}
{\tilde f}_{kn} = \frac{1}{(2\pi)^2}
\int_0^{2\pi} dw^r~ \int_0^{2\pi} dw^\theta~
f[r(w^r),\theta(w^\theta)]
e^{i\left(kw^\theta + nw^r\right)}\;,
\label{eq:ftilde_def}
\end{equation}
and the method for converting the accessible components ${\tilde f}_{kn}$ into the measurable components $f_{kn}$ is described in \cite{Drasco:2004}.
\subsection{Consistency check of the fundamental frequencies}
\label{sec:check_omega}

To confirm the accuracy of our analytical approximations, We compare the analytical results in this paper with that of numerical integration method\cite{zc2022}. Within the analytical framework, the computation of angular frequencies proceeds at a rate substantially higher than that achieved by previous numerical approaches. Specifically, for identical orbital parameters, the analytical model demonstrates at least an order-of-magnitude improvement in computational speed. In Table~\ref{frequency comp}, we evaluate the relative differences of $\Omega_r$, $\Omega_\theta$ and $\Omega_\phi$ between our semi-analytical formalism and numerical integration through five characteristic orbital regimes. It was known that the motion of test particle has a precise analytical solution. For EMRIs where the mass ratio cannot be neglected, we previously derived an analytical EOB orbital solution whose fundamental frequencies were computed using numerical integration methods. The right-most column of this table is the relative frequency shift divided by mass-ratio: $\Delta\Omega/(\Omega\nu)$, where $\Delta \Omega/\Omega$ is the relative difference of radial/polar/azimuthal frequency between the analytical frequency and the numerical ones, i.e., $(\Omega_{\rm ana}-\Omega_{\rm num})/\Omega_{\rm num}$.

\begin{table}[!h]
\begin{center}
\caption{The relative differences of orbital frequencies $\Omega_r$, $\Omega_\theta$ and $\Omega_\phi$ between the analytical and numerical results ($\nu=10^{-3}$). Our analytical results are consistent with numerical results. Relative errors of the orbital frequencies are always less than $10^{-2}\nu$}
{\begin{tabular}{c|c|c|c|c|c|c|c|c}
\hline$a/M$&$p/M$&$e$&$\theta_{min}$&&test-particle& Numerical &Analytical& $\frac{\Delta \Omega}{\Omega}(/\nu)$\\ [5pt]
\hline
\multirow{3}{*} {$0.6$} &\multirow{3}{*} {$8$}&\multirow{3}{*} {$0.6$}&\multirow{3}{*} {$\pi/4$}&$\Omega_r$&$0.01649855$&$0.01650422$&$0.01650428$&$0.004$\\ [3pt]
&&&&$\Omega_\theta$&$0.02557023$&$0.02556221$&$0.02556229$&$0.003$\\[3pt]
&&&&$\Omega_\phi$&$0.02696617$&$0.02695752$&$0.02695762$&$0.004$\\[3pt]\hline
\multirow{3}{*} {$0.6$}
&\multirow{3}{*}{$6$}&\multirow{3}{*}{$0.6$}&\multirow{3}{*} {$\pi/4$}&$\Omega_r$&$0.02050055$&$0.02053345$&$0.02053382$&$0.018$\\[3pt]
&&&& $\Omega_\theta$&$0.04351660$&$0.04344860$&$0.04344908$&$0.011$\\[3pt] 
&&&&$\Omega_\phi$&$0.04756132$&$0.04748323$&$0.04748384$&$0.013$\\[3pt]\hline
\multirow{3}{*} {$0.6$} &\multirow{3}{*} {$6$}&\multirow{3}{*}{$0.9$}&\multirow{3}{*} {$\pi/4$} & $\Omega
_r$&$0.00477440$&$0.00477908$&$0.00477911$&$0.007$\\ [3pt]
&&&&$\Omega_\theta$&$0.01194013$&$0.01184045$&$0.01184048$&$0.003$\\[3pt] 
&&&&$\Omega_\phi$&$0.01339693$&$0.01327828$&$0.01327846$&$0.013$\\[3pt]\hline
\multirow{3}{*} {$0.9$} &\multirow{3}{*} {$6$}&\multirow{3}{*} {$0.9$}&\multirow{3}{*} {$\pi/4$}&$\Omega_r$&$0.00502229$&$0.00502392$&$0.00502395$&$0.007$\\[3pt]
&&&&$\Omega_\theta$&$0.00817252$&$0.00815212$&$0.00815205$&$-0.009$\\[3pt]
&&&&$\Omega_\phi$&$0.00931560$&$0.00929156$&$0.00929164$&$0.009$\\[3pt]\hline
\multirow{3}{*} {$0.9$} &\multirow{3}{*} {$6$}&\multirow{3}{*} {$0.9$}&\multirow{3}{*} {$\pi/3$}&$\Omega_r$&$0.00508130$&$0.00508275$&$0.00508279$&$0.007$\\ [3pt]
&&&&$\Omega_\theta$&$0.00762091$&$0.00760416$&$0.00760405$&$-0.014$\\[3pt]
&&&&$\Omega_\phi$&$0.00863082$&$0.00861138$&$0.00861148$&$0.012$\\[3pt]\hline
\multirow{3}{*} {$0.3$} &\multirow{3}{*} {$8$}&\multirow{3}{*} {$0.6$}&\multirow{3}{*} {$\pi/3$}&$\Omega_r$&$0.01569052$&$0.01569673$&$0.01569691$&$0.011$\\ [3pt]
&&&&$\Omega_\theta$&$0.02708987$&$0.02707981$&$0.02707967$&$-0.005$\\[3pt]
&&&&$\Omega_\phi$&$0.02787234$&$0.02786179$&$0.02786164$&$-0.005$\\
[3pt]\hline
\multirow{3}{*} {$0$} &\multirow{3}{*} {$8$}&\multirow{3}{*} {$0.6$}&\multirow{3}{*} {$\pi/3$}&$\Omega_r$&$0.01408153$&$0.01409080$&$0.01409091$&$0.007$\\ [3pt]
&&&&$\Omega_\theta$&$0.03056697$&$0.03054567$&$0.03054581$&$0.005$\\[3pt]
&&&&$\Omega_\phi$&$0.03056697$&$0.03054567$&$0.03054581$&$0.005$\\
[3pt]\hline
\end{tabular}}
\label{frequency comp}
\end{center}
\end{table}

The frequency shift $\Delta\Omega/(\Omega\nu)$ due to the mass-ratio is almost independent with mass-ratio itself, and the relative shift $\Delta\Omega/\Omega$ is in a range about $[10^{-3}\nu,10^{-2}\nu]$ based on Table \ref{frequency comp}. This indicates that the analytical expressions of the fundamental frequencies in this paper agree with the results of numerical integration method. These facts show that
the our analytical expressions are correct in the cases of generic bound orbits.
\section{Conclusions and Outlook}
\label{sec:conclusion}
Within the EOB's deformed Kerr spacetime, we present the first complete analytical solutions for bound timelike geodesics, expressed through canonical elliptic integrals using Mino time as the evolution parameter. Our analysis shows that the radial and polar motions can be solved in terms of Jacobi elliptic functions, providing significantly greater computational efficiency than numerical integration of the equations of motion. We derive explicit expressions for the three orbital frequencies ($\Omega_r$, $\Omega_\theta$, $\Omega_\phi$) via Fourier series expansion, yielding simpler forms than those in \cite{zc2022}. In the present work, a few approximations have been used. As we mentioned before, in the present model we temporarily omitted the effective spin of the small object. In the EOB theory, this spin of the effective test particle is $\sim \mu a /M$ even if the small object does not rotate. The omission of this term will only induce a relative error of the Hamiltonian at least two orders lower than the mass ratio\cite{zc2021}. Furthermore, The analytical reduction to elliptic integrals is achieved through the approximation $R(r) \approx \mathcal{R}(r)$, with $\mathcal{R}(r)$ being a quartic polynomial that preserves the key orbital turning points. This usually induces an error at $O(10^{-2})\nu$ order, even at the edge of the LSO, the error still $\lesssim 0.1\nu$. The analysis of these approximations performed here showed that the errors could be ignored for EMRIs due to the very small mass ratio  (see Table I). We checked the consistency of the analytical expressions comparing them with the numerical integration method.

Since observer time t is periodic with respect to Mino time, it is straightforward to convert the $\lambda$-Fourier expansion into a t-Fourier expansion\cite{Drasco:2004}. We can apply these solutions to the computation of gravitational waves of EMRIs by the Teukolsky formalism~\cite{Teukolsky:1973ha}. Due to the semi-analytical frequency solutions obtained via Fourier expansion, combining them with the frequency-domain Teukolsky equation proves more convenient and enables efficient numerical waveform generation. In the future, we will use the formalisms in this work to generate the orbital evolution and waveforms for EMRIs with mass-ratio correction, and finalize our EOB-Teukolsky EMRI waveform algorithm.
\acknowledgments
This work is supported by the National Key R\&D
Program of China (Grant No. 2021YFC2203002), and
the National Natural Science Foundation of China (Grants
No. 12173071, No. 12473075). This work made use of the High-Performance Computing Resource in the Core Facility for Advanced Research Computing at Shanghai Astronomical Observatory.

\end{document}